\documentclass{llncs}

\usepackage{syntax}
\usepackage{color,soul}

\usepackage{amsmath}
\usepackage{amssymb}
\usepackage{listings}
\usepackage{proof}
\usepackage{stackengine}
\usepackage{graphicx}
\usepackage{enumitem}
\usepackage{wrapfig}

\begin{document}
\title{Reasoning About TSO Programs Using Reduction and Abstraction}
\author{Ahmed Bouajjani\inst{1} \and Constantin Enea\inst{1} \and Suha Orhun Mutluergil\inst{2} \and Serdar Tasiran\inst{3} }
\institute{IRIF, University Paris Diderot \& CNRS, 
\email{\{abou,cenea\}@irif.fr},
\and Koc University, 
\email{smutluergil@ku.edu.tr}
\and Amazon Web Services, 
\email{tasirans@amazon.com}}
\maketitle
\begin{abstract}
We present a method for proving that a program running under the Total Store Ordering (TSO) memory model is robust, i.e., all its TSO computations are equivalent to computations under the Sequential Consistency (SC) semantics. This method is inspired by Lipton's reduction theory for proving atomicity of concurrent programs. For programs which are not robust, we introduce an abstraction mechanism that allows to construct robust programs over-approximating their TSO semantics. This enables the use of proof methods designed for the SC semantics in proving invariants that hold on the TSO semantics of a non-robust program. These techniques have been evaluated on a large set of benchmarks using the infrastructure provided by CIVL, a generic tool for reasoning about concurrent programs under the SC semantics.
\end{abstract}

\section{Introduction}

A classical memory model for shared-memory concurrency is Sequential Consistency~\cite{lamport79} (SC), where the actions of different threads are interleaved while the program order between actions of each thread is preserved. For performance reasons, modern multiprocessors implement weaker memory models, e.g., Total Store Ordering (TSO)~\cite{DBLP:journals/cacm/SewellSONM10} in x86 machines, which relax the program order. For instance, the main feature of TSO is the write-to-read relaxation, which allows reads to overtake writes. This relaxation reflects the fact that writes are buffered before being flushed non-deterministically to the main memory. 

Nevertheless, most programmers usually assume that memory accesses happen instantaneously and atomically like in the SC memory model. This assumption is safe for data-race free programs~\cite{DBLP:journals/tpds/AdveH93}. However, many programs employing lock-free synchronization are not data-race free, e.g., programs implementing synchronization operations and   libraries implementing concurrent objects. In most cases, these programs are designed to be robust against relaxations, i.e., they admit the same behaviors as if they were run under SC. Memory fences must be included appropriately in programs in order to prevent non-SC behaviors. Getting such programs right is a notoriously difficult and error-prone task. Robustness can also be used as a proof method, that allows to reuse the existing SC verification technology. Invariants of a robust program running under SC are also valid for the TSO executions.
Therefore, the problem of checking robustness of a program against relaxations of a memory model is important. 

In this paper, we address the problem of checking robustness in the case of TSO. We present a methodology for proving robustness which uses the concepts of left/right mover in Lipton's reduction theory~\cite{lipton75}. Intuitively, a program statement is a left (resp., right) mover if it commutes to the left (resp., right) with respect to the statements in the other threads. These concepts have been used by Lipton~\cite{lipton75} to define a program rewriting technique which enlarges the atomic blocks in a given program while preserving the same set of behaviors. In essence, robustness can also be seen as an atomicity problem: every write statement corresponds to two events, inserting the write into the buffer and flushing the write from the buffer to the main memory, which must be proved to happen atomically, one after the other. However, differently from Lipton's reduction theory, the events that must be proved atomic do not correspond syntactically to different statements in the program. This leads to different uses of these concepts which cannot be seen as a direct instantiation of this theory.

Then, following the idea of combining reduction and abstraction introduced in~\cite{elmas09}, we define a program abstraction technique that roughly, makes reads non-deterministic. This technique can be used in two ways. On one side, it can lead to programs that have exactly the same set of reachable configurations as the original program (but these configurations can be reached in different orders), which can be proved robust using the mover theory. This implies that the original program reaches the same set of configurations both under TSO and SC (thus enabling the preservation of SC invariants). 
On the other side, it can lead to over-approximations of the original program which are robust. In this case, any invariant of the over-approximation under SC is also an invariant for the TSO behaviors of the original program.

We tested the applicability of the proposed reduction and abstraction based techniques on an exhaustive benchmark suite containing 34 challenging programs (from \cite{abdulla15} and \cite{bouajjani13}). These techniques were precise enough for proving robustness of 32 of these programs. One program (presented in Figure~\ref{WSQCode}) is not robust, and required abstraction in order to derive a robust over-approximation. There is only one program which cannot be proved robust using our techniques (although it is robust). We believe however that an extension of our abstraction mechanism to atomic read-write instructions will be able to deal with this case. We leave this question for future work.



\section{Overview}
\label{secOverview}
The TSO memory model allows strictly more behaviors than the classic SC memory model: writes are first stored in a thread-local buffer and non-deterministically flushed into the shared memory at a later time (also, the write buffers
 are accessed first when reading a shared variable). However, in practice, many programs are \emph{robust}, i.e., they have exactly the same behaviors under TSO and SC. Robustness implies for instance, that any invariant proved under the SC semantics is also an invariant under the TSO semantics.
We describe in the following a sound methodology for checking that a program is \emph{robust}, which avoids modeling and verifying TSO behaviors.  Moreover, for non-robust programs, we show an abstraction mechanism that allows to obtain robust programs over-approximating the behaviors of the original program.
\begin{figure}
\hspace{1cm}
\begin{minipage}[c]{0.55\textwidth}
\begin{tabular}{l||l}
\begin{lstlisting}
procedure send(){
   y := r1;
   x := 1;
}
\end{lstlisting} \hspace{.2cm}& \hspace{.2cm}
\begin{lstlisting}
procedure recv(){
   do{   
      r1 := x;
   }while(r1 == 0);
   r2 := y;
}
\end{lstlisting} 
\end{tabular}
\end{minipage}
\vline\hspace{5mm}
\begin{minipage}[c]{0.1\textwidth}
 	\includegraphics[trim={6cm 5cm 6cm 3cm},clip, scale=0.2]{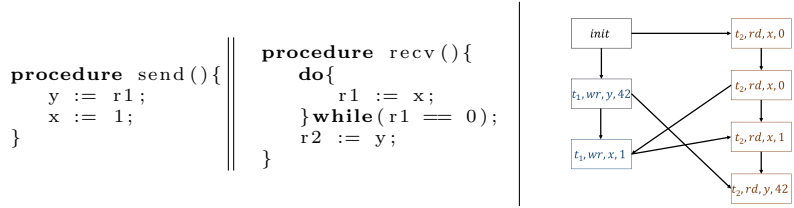}
\end{minipage}

\caption{An example message passing program and a sample trace. Edges of the trace shows the happens before order of global accesses and they are simplified by applying transitive reduction.}
\label{messagePassing}
\end{figure}

As a first example, consider the simple ``message passing'' program in Figure~\ref{messagePassing}.
The \texttt{send} method sets the value of the ``communication'' variable $y$ to some predefined value from register $r1$. Then, it raises a flag by setting the variable $x$ to $1$. Another thread executes the method \texttt{recv} which waits until the flag is set and then, it reads $y$ (and stores the value to register $r2$).
This program is robust, the TSO memory model does not enable new behaviors although the writes may be delayed.
For instance, consider the following TSO execution (we assume that $r1$ contains $42$):
{\footnotesize
\begin{align*}
(t_1,isu)&&\hspace{-4mm}(t_1,isu)(t_1,com,y,42)&&\hspace{-4mm}(t_1,com,x,1) \\
&(t_2,rd,x,0)&&(t_2,rd,x,0)&&(t_2,rd,x,1)(t_2,rd,y,42)
\end{align*}}
The actions of each thread ($t_1$ or $t_2$) are aligned horizontally, they are either \emph{issue} actions ($isu$) for writes being inserted into the thread local buffer (e.g., the first $(t_1,isu)$ represents the write of $y$ being inserted to the buffer), \emph{commit} actions ($com$) for writes being flushed to the main memory (e.g., $(t_1,com,y,42)$ represents the write $y := 42$ being flushed and executed on the shared memory), and \emph{read} actions for reading values of shared variables. Note that every assignment in the program generates two actions, an issue and a commit. The issue action is ``local'', it doesn't enable or disable actions of other threads. 



The above execution can be ``mimicked'' by an SC execution. If we had not performed the $isu$ actions of $t_1$ that early but delayed them until just before their corresponding $com$ actions, we would obtain a valid SC execution of the same program with no need to use store buffers:
{\footnotesize
\begin{align*}
&&\hspace{-4mm}(t_1,wr,y,42)&&\hspace{-4mm}(t_1,wr,x,1) \\
&(t_2,rd,x,0)&&(t_2,rd,x,0)&&(t_2,rd,x,1)(t_2,rd,y,42)
\end{align*}}
Above, consecutive $isu$ and $com$ actions are combined into a single \emph{write} action ($wr$). 
This intuition corresponds to an equivalence relation between TSO executions and SC executions: if both executions contain the same actions on the shared variables (performing the same accesses on the same variables with the same values) and the order of actions on the same variable are the same for both executions, we say that these executions have the same \emph{trace}\cite{shasha88}, or that they are \emph{trace-equivalent}. For instance, both the SC and TSO executions given above have the same trace given in Figure~\ref{messagePassing}. 
The notion of trace is used to formalize robustness for programs running under TSO~\cite{bouajjani13}: a program is called \emph{robust} when every TSO execution has the same trace as an SC execution of the same program. 

Our method for showing robustness is based on proving that every TSO execution can be permuted to a trace-equivalent SC execution (where issue actions are immediately followed by the corresponding commit actions). 
We say that an action $\alpha$ moves right until another action $\beta$ in an execution if we can swap $\alpha$ with every later action until $\beta$ 
while preserving the feasibility of the execution (e.g., not invalidating reads and keeping the actions enabled). We observe that if $\alpha$ moves right until $\beta$ then the execution obtained by moving $\alpha$ just before $\beta$ has the same trace with the initial execution. We also have the dual notion of moves-left with a similar property. As a corollary, if every issue action moves right until the corresponding commit action or every commit action moves left until the corresponding issue action, we can find an equivalent SC execution. For our execution above, the issue actions of the first thread move right until their corresponding $com$ actions. Note that there is a commit action which doesn't move left: moving $(t_1,com,x,1)$ to the left of $(t_2,rd,x,0)$ is not possible since it would disable this read.

In general, issue actions and other thread local actions (e.g. statements using local registers only) move right of other threads' actions. Moreover, issue actions $(t,isu)$ move right of commit actions of the same thread that correspond to writes issued before $(t,isu)$. 
For the message passing program, the issue actions move right until their corresponding commits in all TSO executions since commits cannot be delayed beyond actions of the same thread (for instance reads). Hence, we can safely deduce that the message passing program is robust. However, this reasoning may fail when an assignment is followed by a read of a shared variable in the same thread.

\begin{wrapfigure}{l}{6.7cm}
\centering
\begin{tabular}{l || r}
\begin{lstlisting}
procedure foo(){
   x := 1;
   r1 := z;
   fence
   r2 := y;
}
\end{lstlisting} \hspace{.3cm} & \hspace{.3cm}
\begin{lstlisting}
procedure bar(){
   y := 1;
   fence
   r3:= x;
}
\end{lstlisting} \\
\end{tabular}
\caption{An example store buffering program.}
\label{storeBuffering}
\end{wrapfigure}

Consider the ``store-buffering'' like program in Figure \ref{storeBuffering}. This program is also robust. However, the issue action generated by {\tt x := 1} might not not always move right until the corresponding commit.
Consider the following execution (we assume that the initial value of $z$ is 5):
\begin{align*}
(t_1,isu) & & (t_1,rd,z,5) & & (t_1,com,x,1) & \ldots \\
& (t_2,isu) & & (t_2,com,y,1) (t_2,\tau)(t_2,rd,x,0) & \ldots
\end{align*}

Here, we assumed that $t_1$ executes \texttt{foo} and $t_2$ executes \texttt{bar}. The ${\tt fence}$ instruction generates an action $\tau$. 
The first issue action of $t_1$ cannot be moved to the right until the corresponding commit action since this would violate the program order. Moreover, the corresponding commit action does not move left due to the read action of $t_2$ on $x$ (which would become infeasible). 

The key point here is that a later read action by the same thread ($(t_1,rd,z,5)$) doesn't allow to move the issue action to the right (until the commit). However, this read action moves to the right of the actions of the other threads. So, we can construct an equivalent SC execution by first moving the read action right after the commit $(t_1,com,x,1)$ and then move the issue action right until the commit action.

In general, we can say that an issue $(t,isu)$ of a thread $t$ moves right until the corresponding commit if each read action of $t$ after $(t,isu)$ can move right until the next action of $t$ that follows both the read and the commit. Actually, this property is not required for all such reads. The read actions that follow a fence cannot happen between the issue and the corresponding commit actions. For instance, the last read action of \texttt{foo} cannot happen between the first issue of \texttt{foo} and its corresponding commit action. Such reads that follow a fence are not required to move right.
%
%
In addition, we can omit the right-moves check for the read actions that read from the thread local buffer (see Section~\ref{secRobustness} for more details). 

In brief, our method for checking robustness does the following for every write instruction (assignment to a shared variable): either the commit action of this write moves left or the actions of later read instructions that come before a fence move right in all executions. This semantic condition can be checked using the concept of movers~\cite{DBLP:journals/cacm/Lipton75} as follows: every write instruction is either a left-mover or all the read instructions that come before a fence and can be executed later than the write (in an SC execution) are right-movers. Note that this requires no modeling and verification of TSO executions.

%

For non-robust programs \textcolor{red}{that might reach different configurations than SC executions}, we define an abstraction mechanism that replaces read instructions with ``non-deterministic'' reads that can read more values than the original instructions. The abstracted program has more behaviors than the original one (under both SC and TSO), but it may turn to be robust. When it is robust, we get that any property of its SC semantics holds also for the TSO semantics of the original program.
%
%
%

\begin{figure}[t]
\begin{tabular}{l||r}
{\footnotesize
\begin{lstlisting}
var H,T,items;

procedure steal(){
   local h,t,res;
L1:h := H;
   t := T;
   if(h $\geq$ t)
     return -1;
   res := items[h];
   if( cas(H,h,h+1) )
     return res;
   else 
     goto L1;
}
\end{lstlisting}
}
\hspace{.3cm} & \hspace{.3cm}
{\footnotesize
\begin{lstlisting}
procedure put(var elt){
   local t;
   t := T;
   items[t] := elt;
   T := t+1;
}

procedure take(){
   local h,t,res;
L1:t := T;
   T := t-1;
   h := H;    //havoc(h, h $\leq$ H);
   if( t < h ){
     T := h;
     return -1;
   }
   res := items[t];
   if( t > h )
     return res;
   T := h+1;
   if( cas(H,h,h+1) )
     return task;
   else
     goto L1;
}
\end{lstlisting}
}
\end{tabular}
\caption{Work Stealing Queue.}
\label{WSQCode}
\end{figure}

Consider the work stealing queue implementation in Figure \ref{WSQCode}. A queue is represented with an array ${\tt items}$. Its head and tail indices are stored in the shared variables ${\tt H}$ and ${\tt T}$, respectively.
There are three procedures that can operate on this queue: any number of threads may execute the \texttt{steal} method and remove an element from the head of the queue, and a single unique thread may execute \texttt{put} or \texttt{take} methods nondeterministically. The \texttt{put} method inserts an element at the tail index and the \texttt{take} method removes an element from the tail index.

This program is not robust. Our robustness check fails on this program because the writes of the worker thread (executing the \texttt{put} and \texttt{take} methods) are not left movers and the read from the variable $H$ in the \texttt{take} method is not a right mover. 
This read is not a right mover w.r.t. successful CAS actions of the \texttt{steal} procedure that increment $H$. 

\textcolor{red}{Worse than that, this program might reach to configurations under TSO semantics that are not possible under SC and properties of the SC executions are not satisfied for TSO executions. If there is a single element in the queue and the \texttt{take} method takes it by delaying its writes after some concurrent \texttt{steals}, one of the concurrent \texttt{steals} might also remove this last element. Popping the same element twice is not possible under SC, but it is possible under TSO semantics. However, we can still prove some properties of this program under TSO.}

We apply our abstraction on this instruction \textcolor{red}{of the \texttt{take} method that reads from $H$} such that instead of reading the exact value of $H$, it can read any value less than or equal to the value of $H$. We write this instruction as ${\tt havoc(h,h \leq H)}$ (it assigns to ${\tt h}$ a nondeterministic value satisfying the constraint ${\tt h} \leq {\tt H}$).
Note that this abstraction is sound in the sense that it reaches more states under SC/TSO than the original program.

The resulting program is robust. The statement ${\tt havoc(h,h \leq H)}$ is a right mover w.r.t. successful CAS actions of the stealer threads. Hence, for all the write instructions, the reachable read instructions become right movers and our check succeeds. The abstract program satisfies the specification of an idempotent work stealing queue (elements can be dequeued multiple times) which implies that the original program satisfies this specification as well.


\section{TSO Robustness}
\label{secRobustness}

We present the syntax and the semantics of a simple programming language used to state our results. We define both the TSO and the SC semantics, an abstraction of executions called \emph{trace}~\cite{shasha88} that intuitively, captures the happens-before relation between actions in an execution, and the notion of robustness. 

\smallskip
\noindent
{\bf Syntax.} We consider a simple programming language which is defined in Figure \ref{syntax}. Each program $\mathcal{P}$ has a finite number of shared variables $\overrightarrow{x}$ and a finite number of threads ($\overrightarrow{t}$). Also, each thread $t_i$ has a finite set of local registers ($\overrightarrow{r_i}$) and a start label $l^0_i$. Bodies of the threads are defined as finite sequences of labelled instructions. Each instruction is followed by a \texttt{goto} statement which defines the evolution of the program counter.
Note that multiple instructions can be assigned to the same label which allows us to write non-deterministic programs and multiple \texttt{goto} statements can direct the control to the same label which allows us to mimic imperative constructs like loops and conditionals. An assignment to a shared variable $\langle var \rangle := \langle expr \rangle$ is called a \emph{write instruction}. Also, an instruction of the form  $\langle reg \rangle := \langle var \rangle$ is called a \emph{read instruction}.

\begin{figure}
\begin{grammar}
<prog> ::= "program" <pid> "vars" <var>$^{*}$ <thread>$^{*}$

<thread> ::= "thread" <tid> "regs" <reg>$^{*}$ "init" <label> "begin" <linst>$^{*}$ "end"

<linst> ::= <label>":" <inst>"; goto" <label>";"

<inst> ::= <var> ":=" <expr> \alt <reg> ":=" <expr> \alt <reg> ":=" <var> \alt "fence" \alt <reg> ":= cas"(<var>, <expr>, <expr>)\alt "skip" \alt "assume" <bexpr> 

\end{grammar}
\caption{Syntax of the programs. The star ($^{*}$) indicates zero or more occurrences of the preceding element. $\langle pid\rangle$, $\langle tid\rangle$, $\langle var\rangle$, $\langle reg\rangle$ and $\langle label \rangle$ are elements of their given domains representing the program identifiers, thread identifiers, shared variables, registers and instruction labels, respectively. $\langle expr \rangle$ is an arithmetic expression over $\langle reg \rangle^{*}$. Similarly, $\langle bexpr \rangle$ is a boolean expression over $\langle reg \rangle^{*}$.
}
\label{syntax}
\end{figure}

Instructions of the program can read from or write to shared variables or registers. However, each instruction can access at most one shared variable. We assume that the program $\mathcal{P}$ comes with a set $\mathcal{D}$ that represents the domain of the variables and the registers; and a set of functions $\mathcal{F}$ that allows us to calculate arithmetic and boolean expressions. 

The \texttt{fence} statement empties the buffer of the executing thread.  The \texttt{cas} (compare-and-swap) instruction checks whether the value of its input variable is equal to its second argument. If so, it writes sets third argument as the value of the variable and returns $true$. Otherwise, it returns $false$. In either case, \texttt{cas} empties the buffer immediately after it executes. The \texttt{assume} statement allows us to check conditions. If the boolean expression it contains holds at that state, it behaves like a \texttt{skip}. Otherwise, the execution blocks. Formal description of the instructions are given in Figure~\ref{TSOSemantics}. 

%
%
%
%
%

\smallskip
\noindent
{\bf TSO Semantics.} 
Under the TSO memory model, each thread maintains a local queue to buffer write instructions. 
A state $s$ of the program is a triple of the form $(pc, mem, buf)$. Let $\mathcal{L}$ be the set of available labels in the program $\mathcal{P}$. Then, $pc: \overrightarrow{t} \rightarrow \mathcal{L}$ shows the next instruction to be executed for each thread, $mem: \bigcup_{t_i \in \overrightarrow{t}} \overrightarrow{r_i} \cup \overrightarrow{x} \rightarrow \mathcal{D}$ represents the current values in shared variables and registers and $buf: \overrightarrow{t} \rightarrow (\overrightarrow{x} \times \mathcal{D})^{*}$ represents the contents of the buffers.

There is a special initial state $s_0 = (pc_0, mem_0, buf_0)$. At the beginning, each thread $t_i$ points to its initial label $l_i^0$ i.e., $pc_0(t_i) = l_i^0$. We assume that there is a special default value $0 \in \mathcal{D}$. All the shared variables and registers are initiated as $0$ i.e., $mem_0(x) = 0$ for all $x \in \bigcup_{t_i \in \overrightarrow{t}} \overrightarrow{r_i} \cup \overrightarrow{x}$. Lastly, all the buffers are initially empty i.e., $buf_0(t_i) = \epsilon$ for all $t_i \in \overrightarrow{t}$.

The transition relation $\rightarrow_{TSO}$ between program states is defined in Figure \ref{TSOSemantics}. 
Transitions are labelled by actions. Each action is an element from $\overrightarrow{t} \times (\{\tau, isu\} \cup (\{com, rd \} \times \overrightarrow{x} \times \mathcal{D}))$. 
Actions keep the information about the thread performing the transition and the actual parameters of the reads and the writes to shared variables. We are only interested in accesses to shared variables, therefore, other transitions are labelled with $\tau$ as thread local actions.

\begin{figure}[t]
\scriptsize{
\infer{(pc,mem,buf) \xrightarrow{(t,isu)}_{TSO}(pc', mem, buf[t \rightarrow buf(t) \circ \langle(x,v)\rangle)]}{x := ae(\overrightarrow{r_t}) \in ins(pc(t)) & v = eval(ae(\overrightarrow{r_t})) & x \in \overrightarrow{x}}
\vspace{10pt}
\infer{(pc,mem,buf) \xrightarrow{(t, com, x, v)}_{TSO}(pc, mem, buf[t \rightarrow buf']}{buf(t) = \langle(x,v)\rangle \circ buf' & x \in \overrightarrow{x}}
\vspace{10pt}
\infer{(pc,mem,buf) \xrightarrow{(t,\tau)}_{TSO}(pc', mem[r \rightarrow v], buf)}{r := ae(\overrightarrow{r_t}) \in ins(pc(t)) & v = eval(ae(\overrightarrow{r_t})) & r \in \overrightarrow{r_t}}
\vspace{10pt}
\infer{(pc,mem,buf) \xrightarrow{(t, rd, x, v)}_{TSO}(pc', mem[r \rightarrow v], buf)}{r := x \in ins(pc(t)) & x \in \overrightarrow{x} & v = mem(x) & x \notin varsOfBuf(buf(t)) & r \in \overrightarrow{r_t}}
\vspace{10pt}
\infer{(pc,mem,buf) \xrightarrow{(t, rd, x, v)}_{TSO}(pc', mem[r \rightarrow v], buf)}{r := x \in ins(pc(t)) & x \in \overrightarrow{x} & buf = \alpha \circ \langle (x,v) \rangle \circ \beta & x \notin varsOfBuf(\beta) & r \in \overrightarrow{r_t}}
\vspace{10pt}
\infer{(pc,mem,buf) \xrightarrow{(t, \tau)}_{TSO}(pc', mem, buf)}{\texttt{fence} \in ins(pc(t)) & buf(t) = \epsilon}
\vspace{10pt}
\infer{(pc,mem,buf) \xrightarrow{(t, isu)(t, com, x, v)}_{TSO}(pc', mem[r \rightarrow 1][x \rightarrow v], buf)}{\begin{array}{ccc} r := cas(x, ae_1(\overrightarrow{r_t}), ae_2(\overrightarrow{r_t})) \in ins(pc(t)) & x \in \overrightarrow{x} &  r \in \overrightarrow{r_t} \\ mem(x) = eval(ae_1(\overrightarrow{r_t})) & buf(t) = \epsilon\; & v = eval(ae_2(\overrightarrow{r_t})) \end{array}}
\vspace{10pt}
\infer{(pc,mem,buf) \xrightarrow{(t, rd, x, v)}_{TSO}(pc', mem[r \rightarrow 0], buf)}{\begin{array}{ccc} r := cas(x, ae_1(\overrightarrow{r_t}), ae_2(\overrightarrow{r_t})) \in ins(pc(t)) & x \in \overrightarrow{x} &  r \in \overrightarrow{r_t} \\ mem(x) \neq eval(ae_1(\overrightarrow{r_t})) & buf(t) = \epsilon\; & v = mem(x) \end{array}}
\vspace{10pt}
\infer{(pc,mem,buf) \xrightarrow{(t, \tau)}_{TSO}(pc', mem, buf)}{\texttt{assume}\ be(\overrightarrow{r_t}) \in ins(pc(t)) & eval(be(\overrightarrow{r_t})) = \top}
}
\caption{The TSO Transition Relation. The function $ins$ takes a label $l$ and returns the set of instructions labeled by $l$.
 We always assume that $pc' = pc[t \rightarrow l']$ where $pc(t): inst\ goto\ l';$ is a labelled instruction of $t$ and $inst$ is the instruction described at the beginning of the rule. 
The evaluation function $eval$ calculates the value of an arithmetic or boolean expression based on $mem$ ($ae$ stands for arithmetic expression). Sequence concatenation is denoted by $\circ$. The function $varsOfBuf$ takes a sequence of pairs and returns the set consisting of the first fields of these pairs. 
}
\label{TSOSemantics}
\end{figure}

A TSO execution of a program $\mathcal{P}$ is a sequence of actions $\pi = \pi_1, \pi_2, \ldots, \pi_n$ such that there exists a sequence of states $\sigma = \sigma_0, \sigma_1, \ldots, \sigma_n$, $\sigma_0 = s_0$ is the initial state of $\mathcal{P}$ and $\sigma_{i-1} \xrightarrow{\pi_i} \sigma_i$ is a valid transition for any $i \in \{1,\ldots,n\}$. We assume that buffers are empty at the end of the execution.

\smallskip
\noindent
{\bf SC Semantics.} Under SC, a program state is a pair of the form $(pc, mem)$ where $pc$ and $mem$ are defined as above. Shared variables are read directly from the memory $mem$ and every write updates directly the memory $mem$. To make the relationship between SC and TSO executions more obvious, every write instruction generates $isu$ and $com$ actions which follow one another in the execution (each $isu$ is immediately followed by the corresponding $com$). Since there are no write buffers, \texttt{fence} instructions have no effect under SC. 

\smallskip
\noindent
{\bf Traces and TSO Robustness.} Consider a (TSO or SC) execution $\pi$ of $\mathcal{P}$. The trace of $\pi$ is a graph, denoted by $Tr(\pi)$: Nodes of $Tr(\pi)$ are actions of $\pi$ except the $\tau$ actions. In addition, $isu$ and $com$ actions are unified in a single node. The $isu$ action that puts an element into the buffer and the corresponding $com$ action that drains that element from the buffer correspond to the same node in the trace. 
Edges of $Tr(\pi)$ represent the happens before order ($hb$) between these actions. The $hb$ is union of four relations. The program order $po$ keeps the order of actions performed by the same thread excluding the $com$ actions.  The store order $so$ keeps the order of $com$ actions on the same variable \textcolor{red}{that write different values}. The read-from relation, denoted by $rf$, relates a $com$ action to a $rd$ action that reads its value. Lastly, the from-reads relation $fr$ relates a $rd$ action to a $com$ action that overwrites the value read by $rd$; it is defined as the composition of $rf$ and $so$. 

\textcolor{red}{Our $hb$ relation is slightly different than the standard definition. In the standard definition, two $com$ actions are related by $so$ if they operate on the same variable and write the same value. In our definition, they are not related by $so$. In addition, $fr$ relation implicitly changes due to the difference in $so$. If a $com$ action overwrites a value read by $rd$ but writes the same value, these $rd$ and $com$ actions are not related by $fr$ according to our definition. However, they are related according to the standard definition. This relaxation on $hb$ definition is necessary to relate trace robustness and mover concepts later. In order not to confuse standard and new definitions, we will denote standard traces with $Tr_{std}$.}

We say that the program $\mathcal{P}$ is TSO robust if for any TSO execution $\pi$ of $\mathcal{P}$, there exists an SC execution $\pi'$ such that $Tr(\pi) = Tr(\pi')$. \textcolor{red}{For the standard traces,} it has been proved that robustness implies that the program reaches the same valuations of the shared memory under both TSO and SC~\cite{bouajjani13}. Moreover, the following result characterizes TSO executions that have the same trace as an SC execution.


\begin{lemma}[\cite{shasha88}]
\label{AcyclicTraceLem}
A TSO-execution $\pi$ of a program $\mathcal{P}$ has the same \textcolor{red}{standard} trace as an SC-execution of $\mathcal{P}$ if and only if the happens-before order in \textcolor{red}{ $Tr_{std}(\pi)$} is acyclic.
\end{lemma}

\textcolor{red}{However, Lemma~\ref{AcyclicTraceLem} is not entirely true for the new trace definition. We have shown that only ``only if'' direction holds for the new trace definition and there can be acyclic TSO executions that are not possible under SC.}

\textcolor{red}{Since the new $hb$ definition is a subset of the standard definition, it is easy to see that standard trace-robustness implies new trace-robustness. Moreover, we have shown that new trace-robustness implies both TSO and SC executions reach to the same set of valuations of the shared memory. (I do not provide proofs due to space restrictions.)}


\section{A Reduction Theory for Checking Robustness}

We present a methodology for checking robustness which builds on concepts introduced in Lipton's reduction theory~\cite{DBLP:journals/cacm/Lipton75}. This theory allows to rewrite a given concurrent program (running under SC) into an equivalent one that has larger atomic blocks. Proving robustness is similar in spirit in the sense that one has to prove that issue and commit actions can happen together atomically. However, differently from the original theory, these actions do not correspond to different statements in the program (they are generated by the same write instruction). Nevertheless, we show that the concepts of left/right movers can be also used to prove robustness.

\smallskip
\noindent
{\bf Movers.} Let $\pi = \pi_1,\ldots,\pi_n$ be an SC execution. We say that the action $\pi_i$ \emph{moves right (resp., left)} in $\pi$ if $\pi_1,\ldots,\pi_{i-1}, \pi_{i+1},\pi_i,\pi_{i+2},\ldots,\pi_n$ (resp., $\pi_1,\ldots,\pi_{i-2},\pi_i,\pi_{i-1},\pi_{i+1}\ldots,\pi_n$) is also a valid execution of $\mathcal{P}$, the thread of $\pi_i$ is different than the thread of $\pi_{i+1}$ (resp., $\pi_{i-1}$), and both executions reach to the same end state $\sigma_n$. Since every issue action is followed immediately by the corresponding commit action, an issue action moves right, resp., left, when the commit action also moves right, resp., left, and vice-versa.

Let $\mathsf{instOf}_\pi$ be a function, depending on an execution $\pi$, which given an action $\pi_i \in \pi$, gives the labelled instruction that generated $\pi_i$. Then, a labelled instruction $\ell$ is a \emph{right (resp., left) mover} if for all SC executions $\pi$ of $\mathcal{P}$ and for all actions $\pi_i$ of $\pi$ such that $\mathsf{instOf}(\pi_i) = \ell$, $\pi_i$ moves right (resp., left) in $\pi$. 

A labelled instruction is a \emph{non-mover} if it is neither left nor right mover, and it is a \emph{both mover} if it is both left and right mover.

\smallskip
\noindent
{\bf Reachability Between Instructions.} An instruction $\ell'$ is \emph{reachable from} the instruction $\ell$ if $\ell$ and $\ell'$ both belong to the same thread and there exists an SC execution $\pi$ and indices $1 \leq i<j \leq |\pi|$ such that $\mathsf{instOf}_\pi(\pi_i) = \ell$ and $\mathsf{instOf}_\pi(\pi_j) = \ell'$. We say that $\ell'$ is reachable from $\ell$ \emph{before a fence} if $\pi_k$ is not an action generated by a \texttt{fence} instruction in the same thread as $\ell$, for all $i<k<j$. 
When $\ell$ is a write instruction and $\ell'$ a read instruction, we say that $\ell'$ is \emph{buffer-free} reachable from $\ell$ if $\pi_k$ is not an action generated by a \texttt{fence} instruction in the same thread as $\ell$ or a write action on the same variable that $\ell'$ reads-from, for all $i<k<j$.

\begin{definition}\label{def:atomic}
We say that a write instruction $\ell_w$ is \emph{atomic} if it is a left mover or every read instruction $\ell_r$ buffer-free reachable from $\ell_w$ is a right mover. We say that $\mathcal{P}$ is \emph{write atomic} if every write instruction $\ell_w$ in $\mathcal{P}$ is atomic.
\end{definition}

Note that all of the notions used to define write atomicity (movers and instruction reachability) are based on SC executions of the programs. The following result shows that write atomicity implies robustness. 

\begin{theorem}[Soundness]
\label{soundnessthm}
If $\mathcal{P}$ is write atomic, then it is robust.
\end{theorem}

We will prove the contrapositive of the statement. For the proof, we need the notion of minimal violation defined in \cite{bouajjani13}. A minimal violation is a TSO execution  \textcolor{red}{in which sum of number of same thread actions between $isu$ and corresponding $com$ actions for all writes is minimum. A minimal violation is} of the form $\pi = \pi_1, (t,isu), \pi_2, (t,rd,y,*), \pi_3, (t, com, x, *), \pi_4$ such that $\pi_1$ is an SC execution, only $t$ can delay $com$ actions, the first delayed action is the $(t,com,x,*)$ action after $\pi_3$ and it corresponds to $(t,isu)$ after $\pi_1$, $\pi_2$ does not contain any $com$ or $fence$ actions by $t$ (writes of $t$ are delayed until after $(t,rd,y,*)$), $(t,rd,y,*) \rightarrow_{hb^+} act$ for all $act \in \pi_3 \circ \{(t,com,x,*)\}$ ($isu$ and $com$ actions of other threads are counted as one action for this case), $\pi_3$ does not contain any action of $t$, $\pi_4$ contains only and all of the $com$ actions of $t$ that are delayed in $(t,isu) \circ \pi_2$ and none of the $com$ actions in $(t,com,x,*) \circ \pi_4$ touches $y$.

Minimal violations are important for us because of the following property:
\begin{lemma}[Completeness of Minimal Violations \cite{bouajjani13}] The program $\mathcal{P}$ is robust iff it does not have a minimal violation.
\label{MinimalLemma}
\end{lemma} 

\textcolor{red}{Lemma~\ref{MinimalLemma} defined on standard traces also hold for our extended trace definition.}

Before going into the proof of Theorem~\ref{soundnessthm}, let us define some notation. Let $\pi$ be a sequence representing an execution or a fragment of it. Let $Q$ be a set of thread identifiers. Then, $\pi|_Q$ is the projection of $\pi$ on actions from the threads in $Q$. Similarly, $\pi|_n$ is the projection of $\pi$ on first $n$ elements for some natural number $n$. $sz(\pi)$ gives the length of the sequence $\pi$. We also define a product operator $\otimes$. Let $\pi$ and $\rho$ be some execution fragments. Then, $\pi \otimes \rho$ is same as $\pi$ except that if the $i^{th}$ $isu$ action of $\pi$ is not immediately followed by a $com$ action by the same thread, then $i^{th}$ $com$ action of $\rho$ is inserted after this $isu$. \textcolor{red}{Product operator helps us to fill unfinished writes in one execution fragment by inserting commit actions from another fragment immediately after the issue actions.}

\begin{proof}[Theorem~\ref{soundnessthm}]
Assume $\mathcal{P}$ is not robust. Then, there exists a minimal violation $\pi = \pi_1, \alpha, \pi_2, \theta, \pi_3, \beta, \pi_4$ satisfying the conditions described before, where $\alpha = (t, isu)$, $\theta = (t, rd, y, *)$ and $\beta = (t, com, x, *)$. Below, we show that the write instruction $w = \mathsf{instOf}(\alpha)$ is not atomic.
\begin{enumerate}[label*=\arabic*.]
\item $w$ is not a left mover.
	 \begin{enumerate}[label*=\arabic*.]
	 \item $\rho = \pi_1, \pi_2|_{\overrightarrow{t}\backslash \{t\}}, \pi_3|_{\overrightarrow{t}\backslash \{t\}}|_{sz(\pi_3|_{\overrightarrow{t}\backslash \{t\} })-1}, \gamma, (\alpha, \beta)$ is an SC execution of $\mathcal{P}$ where $\gamma$ is the last action of $\pi_3$. $\gamma$ is a read or write action on $x$ performed by a thread $t'$ other than $t$ and value of $\gamma$ is different from what is written by $\beta$.
	 	\begin{enumerate}[label*=\arabic*.]
	 	\item $\rho$ is an SC execution because $t$ never changes value of a shared variable in $\pi_2$ and $\pi_3$. So, even we remove actions of $t$ in those parts, actions of other threads are still enabled. Since other threads perform only SC operations in $\pi$, $\pi_1, \pi_2|_{\overrightarrow{t}\backslash \{t\}}, \pi_3|_{\overrightarrow{t}\backslash \{t\}}$ is an SC execution. From $\pi$, we also know that the first enabled action of $t$ is $\alpha$ if we delay the actions of $t$ in $\pi_2$ and $\pi_3$.
	 	\item The last action of $\pi_3$ is $\gamma$. By definition of a minimal violation, we know that $\theta \rightarrow_{hb^+} \alpha$ and $\pi_3$ does not contain any action of $t$. So, there must exist an action $\gamma \in \pi_3$ such that either $\gamma$ reads from $x$ and $\gamma \rightarrow_{fr} \beta$ in $\pi$ or $\gamma$ writes to $x$ and $\gamma \rightarrow_{st} \beta$ in $\pi$. Moreover, $\gamma$ is the last action of $\pi_3$ because if there are other actions after $\gamma$,  we can delete them and can obtain another minimal violation which is shorter than $\pi$ and hence contradict the minimality of $\pi$. 
	 	\end{enumerate}
	 \item $\rho' = \pi_1, \pi_2|_{\overrightarrow{t}\backslash \{t\}}, \pi_3|_{\overrightarrow{t}\backslash \{t\}}|_{sz(\pi_3|_{\overrightarrow{t}\backslash \{t\} })-1}, (\alpha, \beta), \gamma$ \textcolor{red}{is an SC execution with a different end state than $\rho$ defined in $1.1$ has or it is not an SC execution, where $\mathsf{instOf}(\gamma') = \mathsf{instOf}(\gamma)$.}
	 	\begin{enumerate}[label*=\arabic*.]
	 	\item In the last state of $\rho$, $x$ has the value written by $\beta$. \textcolor{red}{ If $\gamma$ is a write action on $x$, then $x$ has a different value at the end of $\rho'$ due to the definition of a minimal violation ($\gamma$ and $\beta$ should write different values to have an $so$ edge between them according to the our new $hb$ definition although it is not necessary according to the standard $hb$ definition)}. If $\gamma$ is a read action on $x$, then it does not read the value written by $\beta$ in $\rho$, \textcolor{red}{(again due to the new definition of $so$)}. However, $\gamma$ reads this value in $\rho'$ . Hence, $\rho'$ is not a valid SC execution. 
	 	\end{enumerate}
	 \end{enumerate}
\item There exists a read instruction $r$ buffer-free reachable from $w$ such that $r$ is not a right mover. We will consider two cases: Either there exists a $rd$ action of $t$ on variable $z$ in $\pi_2$ such that there is a later write action by another thread $t'$ on $z$ in $\pi_2$ that writes a different value or not. Moreover, $z$ is not a variable that is touched by the delayed commits in $\pi_4$ i.e., it does not read its value from the buffer.
	\begin{enumerate}[label*=\arabic*.]
		\item We first evaluate the negation of above condition. Assume that for all actions $\gamma$ and $\gamma'$ such that $\gamma$ occurs before $\gamma'$ in $\pi_2$, either $\gamma \neq (t, rd, z, v_z)$ or $\gamma' \neq (t', isu)(t', com, z, v_z')$. Then, $r=\mathsf{instOf}(\theta)$ is not a right mover and it is buffer-free reachable from $w$.  
			\begin{enumerate}[label*=\arabic*.]
			\item $\rho = \pi_1, \pi_2|_{\overrightarrow{t}\backslash \{t\}}, \pi_2|_{\{t\}} \otimes \pi_4, \theta, \theta'$ is a valid SC execution of $\mathcal{P}$ where $\theta'=(t',isu)(t', com, y, *)$ for some $t \neq t'$. 
				\begin{enumerate}[label*=\arabic*.]
				\item $\rho$ is an SC execution. $\pi_1, \pi_2|_{\overrightarrow{t}\backslash \{t\}}$ is a valid SC execution since $t$ does not update value of a shared variable in $\pi_2$. Moreover, all of the actions of $t$ become enabled after this sequence since $t$ never reads value of a variable updated by another thread in $\pi_2$. Lastly, the first action of $\pi_3$ is enabled after this sequence.
				\item The first action of $\pi_3$ is $\theta'=(t',isu)(t', com, y, *)$. Let $\theta'$ be the first action of $\pi_3$. Since $\theta \rightarrow_{hb} \theta'$ in $\pi$ and $\theta'$ is not an action of $t$ by definition of minimal violation, the only case we have is $\theta \rightarrow_{fr} \theta'$. Hence, $\theta'$ is a write action on $y$ \textcolor{red}{that writes a different value than $\theta$ reads.}
				\item $r$ is buffer-free reachable from $w$. $\rho$ is a SC execution, first action of $\rho$ after $\pi_1,\pi_2|_{\overrightarrow{t}\backslash \{t\}}$ is $\alpha, \beta$; $w = \mathsf{instOf}((\alpha,\beta))$, $r = \mathsf{instOf}(\theta)$ and actions of $t$ in $\rho$ between $\alpha, \beta$ and $\theta$ are not instances of a fence instruction or write to $y$.
				\end{enumerate}
			\item $\rho' = \pi_1, \pi_2|_{\overrightarrow{t}\backslash \{t\}}, \pi_2|_{\{t\}} \otimes \pi_4, \theta', \theta$ is not a valid SC execution.
				\begin{enumerate}[label*=\arabic*.]
				\item In the last state of $\rho$, the value of $y$ seen by $t$ is the value read in $\theta$. It is different than the value written by $\theta'$. However, at the last state of $\rho'$, the value of $y$ $t$ sees must be the value $\theta'$ writes. Hence, $\rho'$ is not a valid SC execution.
				\end{enumerate}
			\end{enumerate}
		\item Assume that there exists $\gamma = (t, rd, z, v_z)$ and $\gamma' = (t', isu)(t', com, z,v_z')$ in $\pi_2$. Then, $r= \mathsf{instOf}(\gamma)$ is not a right mover and $r$ is buffer-free reachable from $w$.
			\begin{enumerate}[label*=\arabic*.]
			\item Let $i$ be the index of $\gamma$ and $j$ be the index of $\gamma'$ in $\pi_2$. Then, define $\rho = \pi_1, \pi_2|_{j-1}|_{\overrightarrow{t}\backslash \{t\}}, \pi_2|_i|_{\{t\}} \otimes \pi_4, \gamma'$. $\rho$ is an SC execution of $\mathcal{P}$.
				\begin{enumerate}[label*=\arabic*.]
				\item $\rho$ is an SC execution. $\pi_1, \pi_2|_{j-1}|_{\overrightarrow{t}\backslash \{t\}}$ prefix is a valid SC execution because $t$ does not update any shared variable in $\pi_2$. Moreover, all of the actions of $t$ in $\pi_2|_i|_{\{t\}} \otimes \pi_4$ become enabled after this sequence since $t$ never reads a value of a variable updated by another thread in $\pi_2$ and $\gamma'$ is the next enabled in $\pi_2$ after this sequence since it is a write action.
				\end{enumerate}
			\item Let $i$ and $j$ be indices of $\gamma$ and $\gamma'$ in $\pi_2$ respectively. Define $\rho' = \pi_1, \pi_2|_{j-1}|_{\overrightarrow{t}\backslash \{t\}}, \pi_2|_{i-1}|_{\{t\}} \otimes \pi_4, \gamma', \gamma$. Then, $\rho'$ is not a valid SC execution.
				\begin{enumerate}[label*=\arabic*.]
				\item In the last state of $\rho$, value of $z$ seen by $t$ is $v_z$. It is different than the $v_z'$, value written by $\gamma'$. However, in the last state of $\rho'$, the value of $z$ $t$ sees  must be $v_z'$. Hence, $\rho'$ is not a valid SC execution. 
				\end{enumerate}
			\item $r$ is buffer-free reachable from $w$ because $\rho$ defined in 2.2.1 is an SC execution, first action after $\pi_1, \pi_2|_{j-1}|_{\overrightarrow{t}\backslash \{t\}}$ is $\alpha, \beta$, $w = \mathsf{instOf}((\alpha, \beta))$, $r = \mathsf{instOf}(\gamma)$ and actions of $t$ in $\rho$ between $\alpha, \beta$ and $\theta$ are not instances of a fence instruction or a write to $z$ by $t$.				 
			\end{enumerate}
	\end{enumerate}
\end{enumerate}
\end{proof}

\section{Abstractions and Verifying non-Robust Programs}
\label{secAbstraction}


In this section, we introduce program abstractions which are useful for verifying non-robust TSO programs (or even robust programs -- see an example at the end of this section). In general, a program $\mathcal{P}'$ abstracts another program $\mathcal{P}$ for some semantic model $\mathbb{M}\in \{\text{SC}, \text{TSO}\}$ if every shared variable valuation $\sigma$ reachable from the initial state in an $\mathbb{M}$ execution of $\mathcal{P}$ is also reachable in an $\mathbb{M}$ execution of $\mathcal{P}'$. We denote this abstraction relation as $\mathcal{P} \preceq_\mathbb{M} \mathcal{P}'$.

In particular, we are interested in \emph{read instruction abstractions}, which replace instructions that read from a shared variable with more ``liberal'' read instructions that can read more values (this way, the program may reach more shared variable valuations). We extend the program syntax in Section~\ref{secRobustness} with havoc instructions of the form $\texttt{havoc}(\langle reg \rangle, \langle varbexpr \rangle)$, where $\langle varbexpr \rangle$ is a boolean expression over a set of registers and a single shared variable $\langle var \rangle$. The meaning of this instruction is that the register $reg$ is assigned with any value that satisfies $varbexpr$ (where the other registers and the variable $var$ are interpreted with their current values). 
The program abstraction we consider will replace read instructions of the form $\langle reg \rangle := \langle var \rangle$ with havoc instructions $\texttt{havoc}(\langle reg \rangle, \langle varbexpr \rangle)$.
%

While replacing read instructions with havoc instructions, we must guarantee that the new program reaches at least the same set of shared variable valuations after executing the havoc as the original program after the read. Hence, we allow such a rewriting only when the boolean expression $varbexpr$ is weaker (in a logical sense) than the equality $reg=var$ (hence, there exists an execution of the havoc instruction where $reg = var$). 



\begin{lemma}
\label{lemmaAbs}
Let $\mathcal{P}$ be a program and $\mathcal{P}'$ be obtained from $\mathcal{P}$ by replacing an instruction $l_1: x := r; \texttt{goto}\ l_2$ of a thread $t$ with $l_1: \texttt{havoc}(r,\phi(x,\overrightarrow{r})); \texttt{goto}\ l_2$ such that $\forall x,r.\ x=r \implies  \phi(x,\overrightarrow{r})$ is valid. Then, $\mathcal{P} \preceq_{SC} \mathcal{P}'$ and $\mathcal{P} \preceq_{TSO} \mathcal{P}'$.
\end{lemma}
%

The notion of trace extends to programs that contain havoc instructions as follows. Assume that $(t, hvc, x, \phi(x))$ is the action generated by an instruction $\texttt{havoc}(r,\phi(x,\overrightarrow{r}))$, where $x$ is a shared variable and $\overrightarrow{r}$ a set of registers (the action stores the constraint $\phi$ where the values of the registers are instantiated with their current values -- the shared variable $x$ is the only free variable in $\phi(x)$). Roughly, the $hvc$ actions are special cases of $rd$ actions. Consider an execution $\pi$ where an action $\alpha = (t,hvc, x, \phi(x))$ is generated by reading the value of a write action $\beta = (com, x, v)$ (i.e., the value $v$ was the current value of $x$ when the havoc instruction was executed). Then, the trace of $\pi$ contains a read-from edge $\beta \rightarrow_{rf} \alpha$ as for regular read actions. However, $fr$ edges are created differently. If $\alpha$ was a $rd$ action we would say that we have $\alpha \rightarrow_{fr} \gamma$ if $\beta \rightarrow_{rf} \alpha$ and $\beta \rightarrow_{st} \gamma$. For the havoc case, the situation is a little bit different. Let $\gamma = (com, x, v')$ be an action. We have $\alpha \rightarrow_{fr} \gamma$ if and only if either $\beta \rightarrow_{rf} \alpha$, $\beta \rightarrow_{st} \gamma$ and $\phi(v')$ is false or $\alpha \rightarrow_{fr} \gamma'$ and $\gamma' \rightarrow_{st} \gamma$ where $\gamma'$ is an action. 
Intuitively, there is a from-read dependency from an havoc action to a commit action, only when the commit action invalidates the constraint $\phi(x)$ of the havoc (or if it follows such a commit in store order).

The notion of write-atomicity (Definition~\ref{def:atomic}) extends to programs with havoc instructions by interpreting havoc instructions $\texttt{havoc}(r,\phi(x,\overrightarrow{r}))$ as regular read instructions $r := x$. Theorem~\ref{soundnessthm} which states that write-atomicity implies robustness can also be easily extended to this case.

Read abstractions are useful in two ways. First, they allow us to prove properties of non-robust program as the work stealing queue example in Figure \ref{WSQCode}. We can apply appropriate read abstractions to relax the original program so that it becomes robust in the end. Then, we can use SC reasoning tools on the robust program to prove invariants of the program. 

Second, read abstractions could be helpful for proving robustness directly. The method based on write-atomicity we propose for verifying robustness is sound but not complete. Some incompleteness scenarios can be avoided using read abstractions. If we can abstract read instructions such that the new program reaches exactly the same states (in terms of shared variables) as the original one, it may help to avoid executions that violate mover checks.

\begin{figure}
\begin{tabular}{l || r}
\begin{lstlisting}
procedure foo(){
   x := 1;
   r2 := y;
}
\end{lstlisting} \hspace{.3cm} & \hspace{.3cm}
\begin{lstlisting}[escapeinside={(*}{*)}]
procedure bar(){
   do{   
      r1 = x;
   //havoc(r1,(*$(x \neq 0) ? r1 = x \vee r1 = 0 : r1 = 0$*))
   }while(r1 == 0);
   y := 1;
}
\end{lstlisting} \\
\end{tabular}
\caption{An example program that needs read abstraction to pass our robustness checks. The \texttt{havoc} statement in comments reads as follows: if value of $x$ is not $0$ then $r1$ gets either the value of $x$ or $0$. Otherwise, it is $0$.}
\label{robustButNeedsAbstraction}
\end{figure}

Consider the program in Figure \ref{robustButNeedsAbstraction}. The write statement \texttt{x := 1} in procedure \texttt{foo} is not atomic. It is not a left mover due to the read of $x$ in the do-while loop of \texttt{bar}. Moreover, the later read from $y$ is buffer-free reachable from this write and it is not a right mover because of the write to $y$ in \texttt{bar}. To make it atomic, we apply read abstraction to the read instruction of \texttt{bar} that reads from $x$. In the new relaxed read, $r1$ can read $0$ along with the value of $x$ when $x$ is not zero as shown in the comments below the instruction. With this abstraction, the write to $x$ becomes a left mover because reads from $x$ after the write can now read the old value which was $0$. Consequently, the program becomes write-atomic. \textcolor{red}{If we think of TSO traces of the abstract program and replace $hvc$ nodes with $rd$ nodes, we obtain exactly the TSO traces of the original program. However, the read abstraction adds more SC traces to the program and the program becomes robust.}

\section{Experimental Evaluation}

To test the practical value of our method, we have considered the benchmark for checking TSO robustness described in~\cite{abdulla15}, which consists of 34 programs. This benchmark is quite exhaustive, it includes examples introduced in previous works on this subject. 

Many of the programs in this benchmark are easy to prove being write-atomic. Every write is followed by no buffer-free read instruction which makes them trivially atomic (like the message passing program in Figure \ref{messagePassing}). This holds for 20 out of the 34 programs.

For 13 examples, we needed to perform mover checks and/or read abstractions to show robustness. \textcolor{red}{12 of these examples are robust by our trace-robustness definition but 10 of them are not robust according to the standard trace-robustness. For all 12 programs (except Chase-Lev), both SC and TSO executions reach to the same set of shared variable valuations.} Detailed information for these examples can be found in Table \ref{table:benchmarkResults}.
To check whether writes/reads are left/right movers and the soundness of abstractions, we have used the tool \textsc{Civl}~\cite{hawblitzel15}. This tool allows to prove assertions about concurrent programs (Owicki-Gries annotations) and also to check whether an instruction is a left/right mover.
The buffer-free read instructions reachable from a write before a fence were obtained using a trivial analysis of the control-flow graph (CFG) of the program. 
This method is a sound approximation of the definition in Section 4 but it was sufficient for all the examples.

\begin{table}[t]
\caption{Benchmark results. 
\textcolor{red}{The second column (SV) states whether the original program (without read abstractions) reach to the same shared variable states. SR column shows whether the original program is robust according to the standard definition or not.}
The fourth column (RB) stands for the robustness status of the original program according to our extended $hb$ definition. RA column shows the number of read abstractions performed. RM column represents the number of read instructions that are checked to be right movers and the LM column represents the write instructions that are shown to be left movers. PO shows the total number of proof obligations generated and VT stands for the total verification time in seconds.}
\centering
\begin{tabular}{|c|c|c|c|c|c|c|c|c|}
\hline
Name & SV & SR & RB & RA & RM & LM & PO & VT \\
\hline
Chase-Lev: & - & - & - & 1 & 2 & - & 149 & 0.332 \\
\hline
FIFO-iWSQ: & + & - & + & - & 2 & - & 124 & 0.323 \\
\hline
LIFO-iWSQ: & + & - & + & - & 1 & - & 109 & 0.305\\
\hline
Anchor- iWSQ: & + & - & + & - & 1 & - & 109 & 0.309 \\
\hline
MCSLock: & + & + & + & 2 & 2 & - & 233 & 0.499 \\
\hline
r+detour: & + & - & + & - & 1 & - & 53 & 0.266 \\
\hline
r+detours: & + & - & + & - & 1 & - & 64 & 0.273 \\
\hline
sb+detours+coh: & + & - & + & - & 2 & - & 108 & 0.322 \\
\hline
sb+detours: & + & - & + & - & 1 & 1 & 125 & 0.316 \\
\hline
write+r+coh: & + & - & + & - & 1 & - & 78 & 0.289 \\
\hline
write+r: & + & - & + & - & 1 & - & 48 & 0.261 \\
\hline
dc-locking: & + & + & + & 1 & 4 & 1 & 52 & 0.284 \\
\hline
inline_pgsql: & + & - & + & 2 & 2 & - & 90 & 0.286 \\
\hline
\end{tabular}
\label{table:benchmarkResults}
\end{table}


Our method was not precise enough to prove robustness for only one example, named as \texttt{nbw-w-lr-rl} in \cite{bouajjani13}. This program contains a method with explicit calls to the \texttt{lock} and \texttt{unlock} methods of a spinlock. The instruction that writes to the lock variable inside the \texttt{unlock} method is not atomic, because of the reads from the lock variable and the calls to the \texttt{getAndSet} primitive inside the \texttt{lock} method. Abstracting the reads from the lock variable is not sufficient in this case due to the conflicts with \texttt{getAndSet} actions. However, we believe that read abstractions could be extended to \texttt{getAndSet} instructions (which both read and write to a shared variable atomically) in order to deal with this example.

\section{Related Work}

The weakest correctness criterion that enables SC reasoning for proving invariants of programs running under TSO is \emph{state-robustness} i.e., the reachable set of states is the same under both SC and TSO. However, this problem has high complexity (at least non-primitive recursive for programs with a finite number of threads and a finite data domain~\cite{atig10}). Therefore, it is difficult to come up with an efficient and precise solution. A symbolic decision procedure is presented in \cite{abdulla12} and over-approximate analyses are proposed in \cite{kuperstein11,kuperstein12}.

Due to the high complexity of state-robustness, stronger correctness criteria with lower complexity have been proposed. Trace-robustness (that we call simply robustness in our paper) is one of the most studied criteria in the literature. Bouajjani et al.~\cite{bouajjani11} have proved that deciding trace-robustness is \textsc{PSpace}-complete for a finite number of threads and a finite data domain. 

There are various tools for checking trace-robustness. \textsc{Trencher} (\cite{bouajjani13}) is a sound and complete tool for this purpose that tries to find minimal violations by delaying writes of a single thread. \textsc{Musketeer} (\cite{alglave14}) provides an approximate solution by checking existence of critical cycles on the control-flow graph. Other tools that implement approximate verification procedures have been proposed in ~\cite{alglave11,burckhardt08,burnim11}. Compared to these tools, our approach is either more general, i.e., it allows to prove robustness for programs that have an unbounded number of threads and an unbounded data domain, or it is more precise, than for instance \textsc{Musketeer}. Also, we propose an abstraction mechanism that allows to prove invariants of non-robust programs.


Besides trace-robustness, there are other correctness criteria like triangular race freedom (\textsc{Trf}) and persistence that are stronger than state-robustness. Persistence (\cite{abdulla15}) is incomparable to trace-robustness, and \textsc{Trf} \cite{owens10} is stronger than both trace-robustness and persistence. Our method can verify examples that are state-robust but neither persistent nor \textsc{Trf}.

Reduction and abstraction techniques were used for reasoning on SC programs. \textsc{Qed} (\cite{elmas09}) is a tool that supports statement transformations as a way of abstracting programs combined with a mover analysis. Also, \textsc{Civl} (\cite{hawblitzel15}) allows proving location assertions in the context of the Owicki-Gries logic which is enhanced with Lipton's reduction theory~\cite{lipton75}. Our work enables the use of such tools for reasoning about the TSO semantics of a program.

\bibliographystyle{plain}
\bibliography{biblio}

\begin{thebibliography}{10}

\bibitem{abdulla12}
Parosh~Aziz Abdulla, Mohamed~Faouzi Atig, Yu-Fang Chen, Carl Leonardsson, and
  Ahmed Rezine.
\newblock Counter-example guided fence insertion under tso.
\newblock In {\em International Conference on Tools and Algorithms for the
  Construction and Analysis of Systems}, pages 204--219. Springer, 2012.

\bibitem{abdulla15}
Parosh~Aziz Abdulla, Mohamed~Faouzi Atig, and Tuan-Phong Ngo.
\newblock The best of both worlds: Trading efficiency and optimality in fence
  insertion for tso.
\newblock In {\em European Symposium on Programming Languages and Systems},
  pages 308--332. Springer, 2015.

\bibitem{DBLP:journals/tpds/AdveH93}
Sarita~V. Adve and Mark~D. Hill.
\newblock A unified formalization of four shared-memory models.
\newblock {\em {IEEE} Trans. Parallel Distrib. Syst.}, 4(6):613--624, 1993.

\bibitem{alglave14}
Jade Alglave, Daniel Kroening, Vincent Nimal, and Daniel Poetzl.
\newblock Don’t sit on the fence.
\newblock In {\em International Conference on Computer Aided Verification},
  pages 508--524. Springer, 2014.

\bibitem{alglave11}
Jade Alglave and Luc Maranget.
\newblock Stability in weak memory models.
\newblock In {\em Computer Aided Verification}, pages 50--66. Springer, 2011.

\bibitem{atig10}
Mohamed~Faouzi Atig, Ahmed Bouajjani, Sebastian Burckhardt, and Madanlal
  Musuvathi.
\newblock On the verification problem for weak memory models.
\newblock {\em ACM Sigplan Notices}, 45(1):7--18, 2010.

\bibitem{bouajjani13}
Ahmed Bouajjani, Egor Derevenetc, and Roland Meyer.
\newblock Checking and enforcing robustness against tso.
\newblock In {\em Programming Languages and Systems}, pages 533--553. Springer,
  2013.

\bibitem{bouajjani11}
Ahmed Bouajjani, Roland Meyer, and Eike M{\"o}hlmann.
\newblock Deciding robustness against total store ordering.
\newblock In {\em International Colloquium on Automata, Languages, and
  Programming}, pages 428--440. Springer, 2011.

\bibitem{burckhardt08}
Sebastian Burckhardt and Madanlal Musuvathi.
\newblock Effective program verification for relaxed memory models.
\newblock In {\em International Conference on Computer Aided Verification},
  pages 107--120. Springer, 2008.

\bibitem{burnim11}
Jabob Burnim, Koushik Sen, and Christos Stergiou.
\newblock Sound and complete monitoring of sequential consistency for relaxed
  memory models.
\newblock In {\em International Conference on Tools and Algorithms for the
  Construction and Analysis of Systems}, pages 11--25. Springer, 2011.

\bibitem{elmas09}
Tayfun Elmas, Shaz Qadeer, and Serdar Tasiran.
\newblock A calculus of atomic actions.
\newblock In {\em ACM Symposium on Principles of Programming Languages},
  page~14. Association for Computing Machinery, Inc., January 2009.

\bibitem{hawblitzel15}
Chris Hawblitzel, Shaz Qadeer, and Serdar Tasiran.
\newblock Automated and modular refinement reasoning for concurrent programs.
\newblock {\em Computer Aided Verification}, 2015.

\bibitem{kuperstein11}
Michael Kuperstein, Martin Vechev, and Eran Yahav.
\newblock Partial-coherence abstractions for relaxed memory models.
\newblock In {\em ACM SIGPLAN Notices}, volume~46, pages 187--198. ACM, 2011.

\bibitem{kuperstein12}
Michael Kuperstein, Martin Vechev, and Eran Yahav.
\newblock Automatic inference of memory fences.
\newblock {\em ACM SIGACT News}, 43(2):108--123, 2012.

\bibitem{lamport79}
Leslie Lamport.
\newblock How to make a multiprocessor computer that correctly executes
  multiprocess programs.
\newblock {\em Computers, IEEE Transactions on}, 100(9):690--691, 1979.

\bibitem{lipton75}
Richard~J Lipton.
\newblock Reduction: A method of proving properties of parallel programs.
\newblock {\em Communications of the ACM}, 18(12):717--721, 1975.

\bibitem{DBLP:journals/cacm/Lipton75}
Richard~J. Lipton.
\newblock Reduction: {A} method of proving properties of parallel programs.
\newblock {\em Commun. {ACM}}, 18(12):717--721, 1975.

\bibitem{owens10}
Scott Owens.
\newblock Reasoning about the implementation of concurrency abstractions on
  x86-tso.
\newblock In {\em ECOOP}, volume 6183, pages 478--503. Springer, 2010.

\bibitem{DBLP:journals/cacm/SewellSONM10}
Peter Sewell, Susmit Sarkar, Scott Owens, Francesco~Zappa Nardelli, and
  Magnus~O. Myreen.
\newblock x86-tso: a rigorous and usable programmer's model for x86
  multiprocessors.
\newblock {\em Commun. {ACM}}, 53(7):89--97, 2010.

\bibitem{shasha88}
Dennis Shasha and Marc Snir.
\newblock Efficient and correct execution of parallel programs that share
  memory.
\newblock {\em ACM Transactions on Programming Languages and Systems (TOPLAS)},
  10(2):282--312, 1988.

\end{thebibliography}
\end{document}